\documentclass{mem}
\usepackage{natbib}\usepackage{txfonts}\usepackage{balance}
\usepackage{graphicx}
\usepackage[a4paper,breaklinks,dvipdfm]{hyperref}
\idline{0}{0}
\begin{document}
\def\teff{$T\rm_{eff }$}
\def\kms{$\mathrm {km s}^{-1}$}
\def\msun{\rm\,M_\odot}

\title{Mass segregation in the diffuse outer-halo globular cluster Palomar 14}
\subtitle{}

\author{
Matthias\,J.\,Frank\inst{1,2} 
\and Eva\,K.\,Grebel\inst{1}
\and Andreas\,H.\,W.\,K\"upper\inst{3}}
\offprints{M.\,Frank}
\institute{
Astronomisches Rechen-Institut, Zentrum f\"ur Astronomie der Universit\"at
Heidelberg, M\"onchhofstrasse 12~-~14, D-69120 Heidelberg, Germany
\and
Landessternwarte, Zentrum f\"ur Astronomie der Universit\"at
Heidelberg, K\"onigsstuhl 12, D-69117 Heidelberg, Germany; 
\email{mfrank@lsw.uni-heidelberg.de}
\and
Argelander-Institut f\"ur Astronomie, Auf dem H\"ugel 71, D-53121 Bonn,
Germany
}

\authorrunning{Frank, Grebel \& K\"upper}
\titlerunning{Mass segregation in Palomar 14}

\abstract{We present an analysis of the radial dependence of the stellar mass function in the diffuse outer-halo globular cluster  Palomar\,14. Using archival HST/WFPC2 data of the cluster's central 39\,pc (corresponding to $\sim0.85\times r_h$) we find that the mass function in the mass range $0.55\leq m/\msun\leq 0.85$ is well approximated by a power-law at all radii. The mass function steepens with increasing radius, from a shallow power-law slope of $0.66\pm0.32$ in the cluster's centre to a slope of $1.61\pm0.33$ beyond the core radius, showing that the cluster is mass-segregated. This is seemingly in conflict with its long present-day half-mass relaxation time of $\sim20$\,Gyr, and with the recent finding by \citet{2011ApJ...737L...3B}, who interpret the cluster's non-concentrated population of blue straggler stars as evidence that dynamical segregation has not affected the cluster yet. We discuss this apparent conflict and argue that the cluster must have either formed with primordial mass segregation, or that its relaxation time scale must have been much smaller in the past, i.e. that the cluster must have undergone a significant expansion.

\keywords{Galaxy: globular clusters -- Globular clusters: individual: Palomar 14 -- Galaxies: stellar dynamics -- Stars: formation -- Galaxy: halo}
}
\maketitle{}

\section{Introduction}
Almost all Galactic globular clusters (GC) have present-day half-mass relaxation times shorter than their ages \citep[e.g.][2010 edition]{1996AJ....112.1487H}. In these clusters, two-body relaxation has already altered the distribution of stars: massive stars, losing kinetic energy to lower-mass stars sink into the cluster's centre, whereas low-mass stars gain energy allowing them to populate orbits further away from the cluster's centre. This is observed as mass segregation, i.e. more massive stars show a more concentrated radial distribution than lower-mass stars, and the cluster appears depleted in low-mass stars. 

One of the few exceptions, with a present-day half-mass relaxation time exceeding the Hubble time, is the outer-halo GC Palomar~14 (Pal\,14). According to \citet{2011ApJ...726...47S}, Pal\,14 has a projected half-light radius of $r_h=46$\,pc, making it the most extended Galactic GC in the Milky Way. Its low mass and large radius implies a half-mass relaxation time of $\sim$20 Gyr. Therefore, no  mass segregation is intuitively expected in Pal\,14. In agreement with this expectation, \citet{2011ApJ...737L...3B} found that the cluster's population of blue straggler stars (BSS) is not centrally concentrated compared to red giant (RGB) and horizontal branch (HB) stars. BSS in such a diffuse cluster have most likely formed via mass-transfer in primordial binary systems that have larger total masses than individual RGB or HB stars. Hence, these systems would segregate most quickly in a GC, such that BSS are expected to trace this segregation process. Beccari et al.~interpret their findings as evidence that two-body relaxation has not affected Pal\,14 yet. 

On the other hand, \citet{2009AJ....137.4586J} found that the mass function of main sequence stars in Pal\,14 is described by a power-law $dN/dm\propto m^{-\alpha}$ with a slope of $\alpha=1.3\pm0.4$, i.e. the cluster is significantly depleted in low-mass stars compared to a \citet{2001MNRAS.322..231K} initial mass function (IMF; $\alpha=2.3$ in this mass range). This would readily be understood, if the cluster \emph{were} mass-segregated, or alternatively, if it formed with a IMF already depleted in low-mass stars \citep{2011MNRAS.411.1989Z}.

In this contribution we present evidence that Pal\,14 is mass-segregated based on an analysis of radial dependence of the cluster's stellar mass function (Section~\ref{sec:analysis}) and discuss two possible scenarios that can reconcile our results with the cluster's large relaxation time scale and its non-segregated population of BSS (Section~\ref{sec:discussion}).

\section{Data and Analysis}
\label{sec:analysis}
\begin{figure}[t!]
\includegraphics[width=\linewidth]{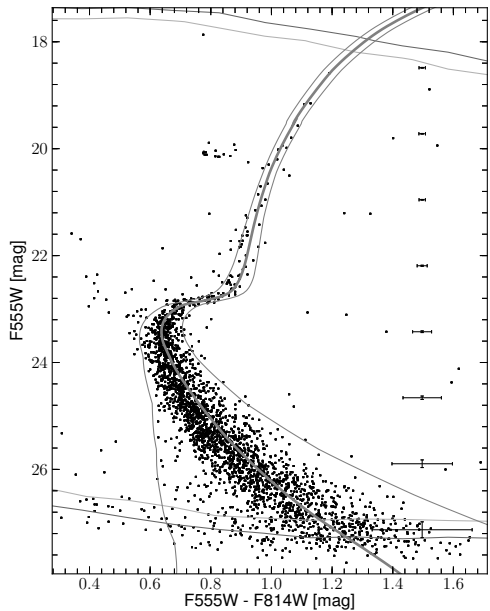}
\caption{\footnotesize
Observed CMD of Pal\,14 obtained from archival WFPC2 data using the \textsc{HSTPHOT} photometry package \citep{2000PASP..112.1383D}. Artificial star tests were used to estimate photometric uncertainties (error bars on the right) and completeness limits (light and dark grey lines at the faint and bright ends correspond to 80\% and 50\% completeness contours, respectively). The isochrone (thick grey curve) corresponds to an age of 11.5 Gyr, [Fe/H]$=-1.5$\,dex and [$\alpha$/Fe]$=+0.2$\,dex assuming a distance of $71\pm1.3$\,kpc and reddening of $E(\mathrm{F555W}-\mathrm{F814W})=0.06$\,mag \citep{2009AJ....137.4586J}. To calculate the mass function we used stars within the colour limits represented by thin grey curves on both sides of the isochrone.}
\label{fig:cmd}
\end{figure}
\begin{figure}[t!]
\includegraphics[width=\linewidth]{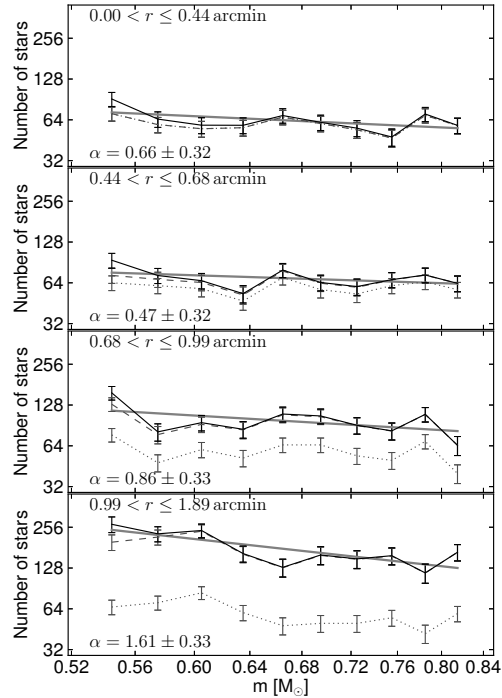}
\caption{\footnotesize
The mass function in four radial ranges in order of increasing distance from the cluster centre. Dotted lines correspond to the raw observed star counts, dashed lines to the star counts after correction for geometric coverage and solid lines to the star counts additionally corrected for photometric incompleteness. The best-fitting power-law mass functions are shown as thick grey lines and the best-fitting slopes $\alpha$ are reported the bottom of each panel. In all radial ranges the mass function is well described by a power-law.}
\label{fig:masssegregation}
\end{figure}
\begin{figure}[t!]
\includegraphics[width=\linewidth]{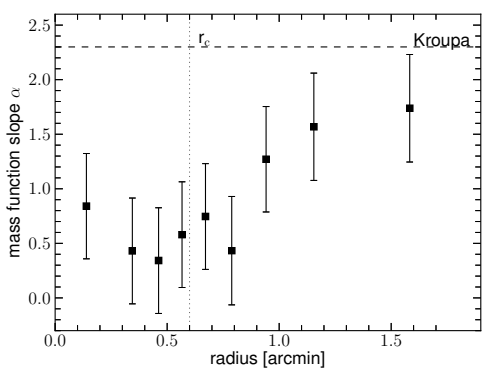}
\caption{\footnotesize
The best-fitting mass function slope and its uncertainties as a function of radius. The core radius $r_c$ of the cluster is indicated by the dotted line. A trend of increasing $\alpha$ with increasing radius is obvious and is significant at the 98\% level.}
\label{fig:masssegregation_alpha_vs_radius}
\end{figure}

We used deep $V$ and $I$ band archival HST/Wide-Field Planetary Camera 2 (WFPC2) imaging of Pal\,14 (program GO 6512, PI: Hesser) to obtain the colour-magnitude diagram (CMD) shown in Fig.~\ref{fig:cmd}. The overlaid isochrone is taken from the \citet{2008ApJS..178...89D} library and corresponds to an age of 11.5 Gyr, [Fe/H]\,$=-1.5$\,dex and [$\alpha$/Fe]$=+0.2$\,dex. To derive the stellar mass function, we followed the basic procedure described in \citet{2012MNRAS.423.2917F}: we selected stars within the colour limits shown as thin grey lines in the CMD and interpolated the masses tabulated in the isochrone file to the observed magnitudes of these stars in order to infer their masses. We corrected for the radial variation of the photometric completeness, as well as the inhomogeneous coverage of the cluster by the WFPC2 pointing, and calculated the maximum likelihood power-law representation of the mass function in different radial ranges. The cluster's observed mass function in radial bins containing each one fourth of the observed stars is shown in Fig.~\ref{fig:masssegregation}. The dotted curves correspond to raw star counts in ten evenly spaced mass bins from 0.54 to 0.82 M$_\odot$, corresponding to the $\sim70$\% completeness limit at the faint end to the tip of the RGB. Dashed curves represent the star counts after correction for geometric coverage, solid curves after additionally correcting for photometric completeness. Thick grey lines show the best-fitting power-law. The mass function is well described by a single power law at all radii and a trend of an increasing power law slope with increasing radius is apparent. This trend is seen more clearly in the finer radial subdivision of Fig.~\ref{fig:masssegregation_alpha_vs_radius}, which shows the best-fitting mass function slope $\alpha$ as a function of radius. The mass function steepens with increasing radius, ranging from $\alpha<1$ within the cluster's core radius to a slope almost compatible with the Kroupa $\alpha=2.3$ in the outermost radial bin at $r=1.6$\,arcmin or 33\,pc. A constant mass function slope as a function of radius is excluded at the 98\% confidence level. The lack of low-mass stars in the cluster's centre compared to larger radii clearly shows that the cluster is mass-segregated.
 
\section{Discussion}
\label{sec:discussion}
Finding mass-segregation in Pal\,14 is in apparent conflict with its present-day half-mass relaxation time of $\sim20$\,Gyr. If the cluster had a similar structure and therefore a similar relaxation time-scale throughout its lifetime, the observed mass segregation would have to be primordial, as was suggested by \citet{2011MNRAS.411.1989Z}. In this case, it is likely that the cluster spent most of its lifetime in the low-density environment of an only recently accreted dwarf galaxy \citep[cf.][]{2011ApJ...726...47S,2012A&A...537A..83C}, or otherwise it is puzzling how such a diffuse cluster can have survived in the tidal field of the Galaxy.  That the cluster is affected by the Galactic tidal field, even at its current remote location (at a Galactocentric distance of 66\,kpc), is evidenced by its tidal tails \citep{2010A&A...522A..71J,2011ApJ...726...47S}.

Alternatively, the cluster may have been significantly more compact (by a factor of $\sim$2 in the projected half-light radius $r_h$) in the past, implying a previously much shorter relaxation time scale $t_{rh}$ of a few Gyrs \citep[$t_{rh}\propto r_{h}^{3/2}$;][]{1971ApJ...164..399S}. Such an expansion could have been caused by tidal shocks during pericenter passages of the cluster on its orbit about the Galaxy, similar to the expansion of Palomar\,5 due to disk shocks that has been suggested by \citep{2004AJ....127.2753D}. Given its present Galactocentric distance of 66\,kpc, Pal\,14 would have to be on a highly eccentric orbit in order to come sufficiently close to the Galactic centre to be affected by tidal shocks. This scenario could not only explain the observed mass segregation, but also the cluster's large physical size, whose light profile in this case may be significantly inflated by unbound stars \citep{2010MNRAS.407.2241K}.

Regarding the non-segregated population of blue stragglers compared to HB and RGB stars, our data confirm the findings of \citep{2011ApJ...737L...3B}: the radial distribution of BSS is not statistically different from that of HB or RGB stars \citep{Franketalinprep}. While the small number statistics ($\sim25$ blue stragglers) advise caution in the interpretation, both of the evolutionary scenarios for Pal\,14 sketched above can potentially be reconciled with a non-segregated population of BSS. A plausible mechanism leading to primordial mass segregation is the `competitive accretion' scenario \citep{2001MNRAS.323..785B}, in which protostars that reside in the cluster's centre, where the density of gas is higher, can accrete gas more efficiently, and consequently tend to have higher masses. In this picture, it seems conceivable that stars are generally segregated by mass, but that the distribution of binaries such as the BSS progenitors is not necessarily more centrally concentrated. If on the other hand the cluster was once significantly more compact, it is possible that not all BSS originate from mass-transfer in primordial binaries, but that a fraction of them formed in collisions in the -- then denser -- cluster centre. If three or more stars were involved in these close encounters (e.g. two binary systems) the resulting BSS would have received initial velocity kicks and would have been expelled from the cluster centre, resulting in a more extended radial distribution of BSS in the cluster. 
 
\begin{acknowledgements}
This work was partially supported by Sonderforschungsbereich 881, ``The Milky Way System'' (subprojects A2 and A3) of the German Research Foundation (DFG) at the University of Heidelberg. M.J.F and A.H.W.K. kindly acknowledge support from the DFG via Emmy Noether Grant Ko 4161/1 and project KR 1635/28-1, respectively. 

\end{acknowledgements}

\bibliographystyle{aa}

\end{document}